\documentclass[aps,prl,twocolumn,nofootinbib,groupedaddress,amsfonts,floatfix]{revtex4}

\usepackage{graphicx,amsmath,amssymb,amstext}
\usepackage{amssymb,amsbsy,amsfonts,amsthm}

\newcommand{\gsim}{\;\mbox{\raisebox{-0.5ex}{$\stackrel{>}{\scriptstyle{\sim}}$}}\;}
\newcommand{\lsim}{\;\mbox{\raisebox{-0.5ex}{$\stackrel{<}{\scriptstyle{\sim}}$}}\;}

\def\bbox{\,\fbox{$\,$}\,}
\def\Bbox{\raisebox{0.1cm}\bbox}

\begin{document}

\title{A New Mechanism for Bubble Nucleation: Classical Transitions}
 
\author{Richard Easther${}^1$}
\author{John T. Giblin, Jr${}^{2,3}$}
\author{Lam Hui${}^{4}$}
\author{Eugene A. Lim${}^{4}$}

\affiliation{${}^1$Department of Physics, Yale University, New Haven CT 06520}   
\affiliation{${}^2$Department of Physics, Kenyon College, Gambier, OH 43022}
\affiliation{${}^3$Perimeter Institute for Theoretical Physics, 31 Caroline St N, Waterloo, ON N2L 2Y5}
\affiliation{${}^4$ISCAP and Physics Department, Columbia University, New York, 10027}

\begin{abstract}
Given a scalar field with metastable minima, bubbles  nucleate quantum mechanically.  When bubbles collide, energy stored in the bubble walls is converted into kinetic energy of the field.  This kinetic energy can facilitate the classical nucleation of new bubbles in minima that lie below those of the ``parent'' bubbles.   This process is efficient and classical, and changes the dynamics and statistics of bubble formation in models with multiple vacua, relative to that derived from quantum tunneling.  

\end{abstract}

\maketitle

Consider a potential $V(\phi)$ with many metastable minima with positive vacuum energy.  A typical region in the universe undergoes de Sitter expansion, with Hubble parameter $H\sim \sqrt{V(\phi_1)}/M_p$, where $M_p$ is the usual reduced Planck mass and $\phi_1$ labels the vacuum.   Small regions may tunnel to another minimum, $\phi_2$, $V(\phi_2) < V(\phi_1)$  \cite{Coleman:1980aw}, forming an expanding ``bubble''.  Some of these bubbles inevitably collide.  In this Letter we show that these collisions can give the field  sufficient energy to climb over potential barrier(s) and form new bubbles at minima where $V(\phi)$ is less than that of either of the original bubbles. Bubble collisions can thus yield new bubble universes -- the progeny become the protagonists. In a universe with many metastable vacua this new {\em classical\/} mechanism populates the de Sitter sea with additional bubbles, beyond those provided by tunneling.

Such classical transitions should not be surprising. Hawking, Moss and Stewart \cite{Hawking:1982ga} numerically analyzed collisions in a potential with two minima.  When cosmological bubbles of the lower vacuum collide, they can form a new bubble of the higher, metastable vacuum. This new bubble is surrounded by regions in the lower vacuum state and thus quickly collapses.  Similar features are seen in  numerical simulations
by \cite{Kosowsky:1991ua,Aguirre:2009}.  The key new result here is that,  given a third, lower energy, minimum, the field can ``slosh'' into that vacuum after a collision between bubbles in higher energy minima. This yields a new bubble which is stable and expanding, since its interior energy density is lower than that of the surrounding region. 

Physically, this process is easy to understand, even though the detailed dynamics require numerical simulations.  A bubble collects energy as it expands into a region of higher potential energy,  storing it  as the gradient energy of its walls. When two bubbles collide, this energy must go somewhere.  The bubble wall may lose coherence and radiate the energy into the space in which the two bubbles have merged \cite{Langlois:2001uq}.  Alternatively, the energy in the walls can be deposited into kinetic energy (in field space), allowing the field to climb potential barriers and thus slosh into adjacent minima. If one of these minima has a lower potential energy than the parent bubbles, we find that this process is efficient, provided the parent bubbles have had a (small) amount of time to expand since they were nucleated, allowing them to store sufficient energy in their walls. Clearly, the transition condition depends on the size of the barriers in the model, relative to the energy stored in the walls.  After nucleation, the collision energy is deposited into the new bubble wall, hence the new bubble is formed with non-stationary walls. 

\mbox{}

\begin{figure}[tbp] 
   \centering
   \includegraphics[width=3in,height=2.2in]{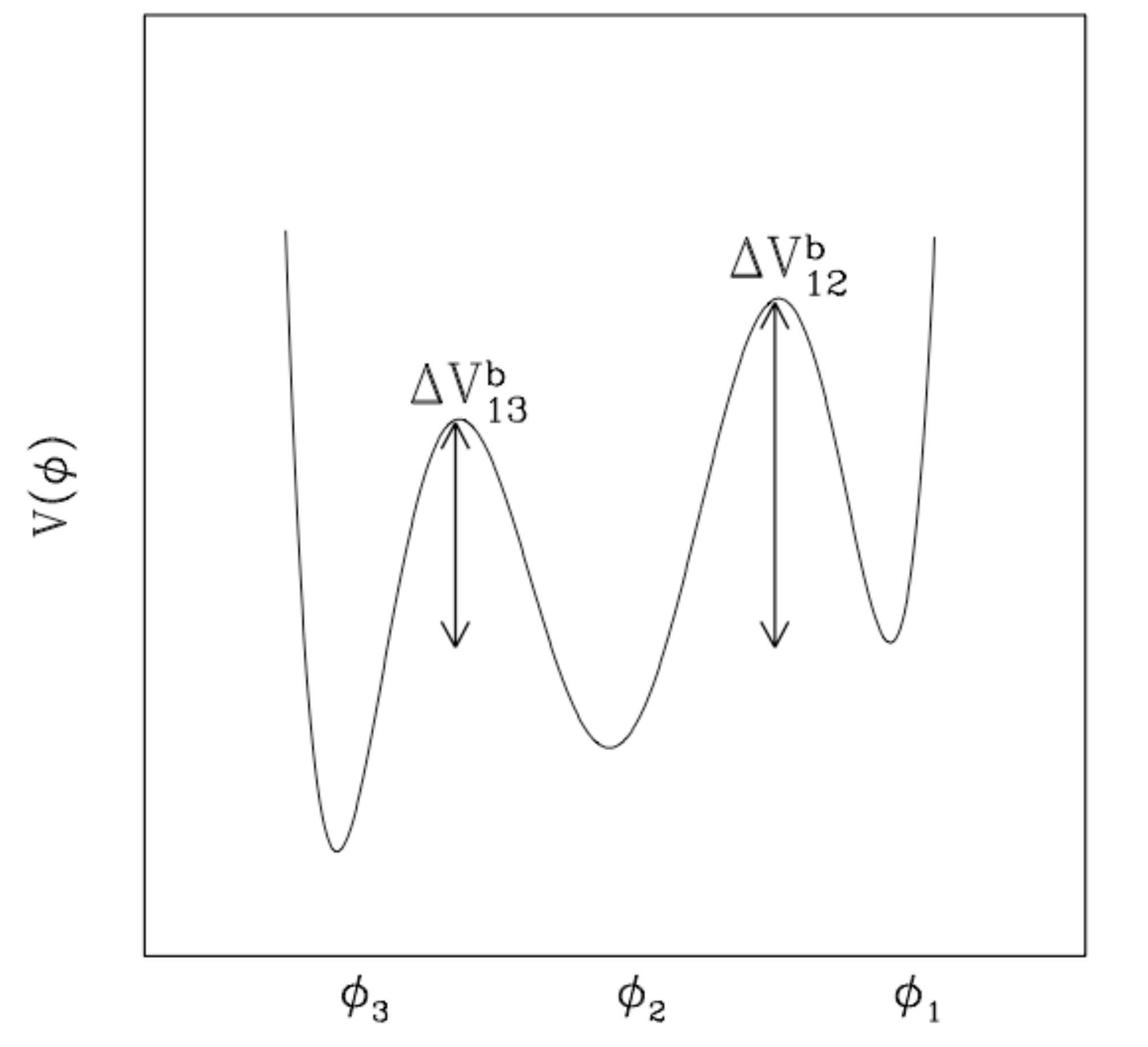} 
   \caption{Schematic diagram of a 3-minima potential. The false vacuum at $\phi_1$ quantum mechanically nucleates bubbles into regions of $\phi_2$. Sufficiently energetic collisions of two $\phi_2$-bubbles can classically nucleate a bubble of $\phi_3$. The two barrier heights are defined with respect to $V(\phi_1)$. Superscript 'b' means barrier.}
   \label{fig:potential}
\end{figure}

\noindent {\em Transition Condition:}   Assume a setup illustrated in Fig. \ref{fig:potential},
with three local minima $\phi_1, \phi_2, \phi_3$.
Two bubbles of $\phi_2$ are quantum mechanically
nucleated within the false vacuum of $\phi_1$. Subsequently, the two $\phi_2$-bubbles 
collide, and the question is: under what condition would the collision trigger the classical
nucleation of a third bubble of $\phi_3$?
For an analytic estimate, we will work in the flat space limit.
The hyperbolic foliation is
convenient for studying collisions \cite{Chang:2007eq,Aguirre:2007an,Aguirre:2007anwm}:
\begin{equation}
ds^2 = -dt^2 + dx^2 + t^2(d\psi^2 + \sinh \psi^2 d\theta^2). \label{eqn:metric}
\end{equation}
The two $\phi_2$-bubbles are nucleated at $t=0$, $x=\pm b$ (and $\psi = 0$).
The pre-collision evolution is well-described by the solution of \cite{Coleman:1980aw}. 
We make the thin-wall approximation, where each bubble wall moves according to
$R_0^2 = (x\pm b)^2 - t^2$, with $R_0$ being the initial bubble radius.
The wall thickness at a given $t$ follows $\Delta x = \gamma^{-1} \Delta x_0$,
where $\Delta x_0$ is the initial thickness, and $\gamma$ is the Lorentz factor:
\begin{eqnarray}
\gamma = {|x \pm b| \over R_0} = \sqrt{1 + {t^2 \over R_0^2}}. \label{eqn:gamma}
\end{eqnarray}
Assuming $b \gg R_0$, the collision occurs at $x=0$ where the
Lorentz boost is $\gamma = b/R_0$ (hereafter $\gamma$ takes this
value). The initial wall thickness can be estimated from the Compton wavelength i.e. 
$\Delta x_0 \sim \Delta \phi_{12} / \sqrt{\Delta V^b_{12}}$, where
$\Delta \phi_{12} \equiv \phi_1 - \phi_2$ and $\Delta V^b_{12}$ is the
barrier in between.

\begin{figure*}[tbp] 
   \centering
   \includegraphics[width= 6.7in]{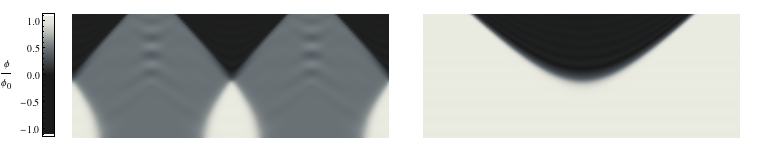}    
   \caption{Conformal diagram showing the values of $\phi$ on the $X-\tau$ plane (left) and the $Y-\tau$ plane (right).  Our conformal coordinates, $(\tau, X, Y, Z)$, are chosen such that $X$ is the axis through the centers of the bubbles and $\tau$ is defined in the caption of Fig \ref{fig:notransition}.  The collision occurs at $X=0$ (analogous to $x=0$ in the text).  Regions of $\phi_1$ appear off-white, regions of $\phi_2$ appear grey and regions of $\phi_3$ appear black.  In this simulation, the centers of the two bubbles are $\sim 0.45 H_0^{-1}$ apart at nucleation.  When these bubbles collide (at $\tau\approx 17.5$) the walls have a Lorentz factor $\gamma \approx 1.5$. The slight ripples in the middle of each bubble are numerical artifacts. }
   \label{fig:conform}
\end{figure*}

The post-collision evolution follows from 
$\Bbox \phi = \partial_{\phi} V$, which in the coordinates of Eqn. (\ref{eqn:metric}) is
\begin{equation}
\frac{\partial^2 \phi}{\partial t^2}+\frac{2}{t}\frac{\partial \phi}{\partial t} = \frac{\partial^2 \phi}{\partial x^2} -\frac{\partial V}{\partial \phi} \label{eqn:EOM}
\end{equation}
Let us focus on the time evolution of $\phi$ at $x = 0$ (and implicitly $\psi = 0$; see Fig. \ref{fig:conform}). 
Pre-collision, $\phi$ is stuck at the false vacuum of $\phi_1$,
where $\partial V/\partial\phi = 0$, and both spatial and time derivatives of $\phi$
vanish. As the collision begins, $\phi$ is still at $\phi_1$, hence
the potential force $\partial V/\partial\phi$ remains zero, but the presence of the two
bubble walls means $\partial^2 \phi /\partial x^2$ is non-zero and gives rise to
a kick towards $\phi_2$ (i.e. negative in the convention of Fig. \ref{fig:potential}):
\begin{eqnarray}
{\partial^2 \phi \over \partial x^2} \sim 
- \gamma^2 {\Delta V^b_{12} \over \Delta \phi_{12}}, \label{eqn:kick}
\end{eqnarray}
where we have used the $1$-bubble results mentioned earlier.
This kick initiates a field motion from $\phi_1$ tending towards $\phi_2$.
Whether this motion can be completed depends on whether the kick wins over
the counteracting potential force ($\sim {\Delta V^b_{12} /\Delta \phi_{12}}$) 
that inevitably develops.
Absent dissipation, to which we will return below,
Eqn. (\ref{eqn:kick}) suggests the kick always wins since we expect $\gamma \gsim 1$
in general. 

Once the field makes it over the first barrier between vacua $1$ and $2$,
it would appear the field can also overcome the second barrier between vacua $2$
and $3$, {\it if} the second barrier is smaller than the first (absent
dissipation). If not, Eqn. (\ref{eqn:EOM}) and simple energy
conservation suggests the required condition for overcoming the second
barrier is $\gamma^2 \gsim \Delta V^b_{13}/\Delta V^b_{12}$ (see Fig. \ref{fig:potential}
for definitions). Summarizing, the condition for a collision induced classical
transition from vacuum $1$ to vacuum $3$ is
\begin{eqnarray}
\gamma^2 \gsim {1\over 1-\beta} {\,\rm max}\left(1, {\Delta V^b_{13} \over \Delta V^b_{12}}\right). \label{eqn:condition2}
\end{eqnarray}
Here, we have introduced a factor of $1/(1-\beta)$ to approximate the effect of dissipation,
where $\beta$ can be thought of as the fractional energy dissipation.
There could be many sources of dissipation: the second term on the left
of Eqn. (\ref{eqn:EOM}) is one example; other examples include radiation into
fields $\phi$ is coupled to (including itself), gravitational waves (if there are significant deviations
from spherical symmetry), and Hubble friction. 
The last item is simple to estimate: Hubble friction becomes important
if $\Delta \phi_{13} \gsim H^{-1} \partial\phi/\partial t
\sim H^{-1} \gamma \Delta \phi_{12}/\Delta x_0$, where
$\Delta \phi_{13} \equiv \phi_1 - \phi_3$. 
Since the initial bubble wall thickness $\Delta x_0$ is generally a small fraction of $H^{-1}$,
we expect Hubble friction to be unimportant unless $\Delta \phi_{13} \gg \gamma \Delta \phi_{12}$.

If Eqn. (\ref{eqn:condition2}) is satisfied, the (model-dependent) dissipation likely causes 
the field at $x = 0$ to eventually settle in $\phi_3$, that is, unless there
are additional lower minima to the left, in which case further field excursions
are possible. It is interesting to note that Eqn. (\ref{eqn:condition2}) can always be satisfied
if the collision is sufficiently relativistic, i.e. 
if enough time elapses between nucleation and collision. 
If $\beta$ is not too close to unity,
and if $\Delta V^b_{12} \sim \Delta V^b_{13}$, {\it classical transitions over potential
barriers are generic even if the collision is barely relativistic.}
On the other hand, if Eqn. (\ref{eqn:condition2}) is not satisfied, the two bubbles merge and the energy stored in the walls dissipate via one or more of the dissipation mechanisms outlined above.

We numerically test these ideas using a toy model:
\begin{equation}
V(\phi) = \frac{\lambda}{4}\phi^2\left(\phi^2-\phi_0^2\right)^2 + \epsilon \lambda \phi_0^5(\phi + \phi_0) + \alpha \lambda \phi_0^6.
\end{equation}
We assume $\epsilon$ is small enough so that we have three classically stable points, $\phi_1 \approx \phi_0$, $\phi_2\approx 0$ and $\phi_3\approx -\phi_0$.  The barriers are parametrized by $\phi_0$ and $\lambda$, while $\alpha$ sets the overall vacuum energy scale which we assume is much larger than the difference in energy between the minima.  

These minima are perturbatively non-degenerate, and we can use the analytic solution of \cite{Coleman:1980aw} as an approximate solution to tunneling events. 
Two bubbles of $\phi=\phi_2$ nucleate within a sea of highest metastable vacuum $\phi = \phi_1$,
with an initial radius $R_0 = (\sqrt{32\lambda}\epsilon  \phi_0^2/3)^{-1}$.
These bubbles have an approximate initial profile,
\begin{equation}
\phi (r) = \phi_0 \left(1+2e^{-\sqrt{2\lambda}\phi_0^2(r-R_0)}\right)^{-1/2},
\end{equation}
where $r$ is the distance from center of the bubble. The approximate height of the barrier between the top two minima is 
\begin{equation}
\Delta V_{12}^b\approx \frac{\lambda \phi_0^6}{27},
\end{equation}
which is also roughly the barrier between the middle and lower minima $\Delta V_{13}^b \approx \Delta V_{12}^b$.
We explore the transition condition numerically in three spatial dimensions, using a modified version of {\sc LatticeEasy} \cite{Felder:2000hq} and  $1024^3$ lattices. We take the overall vacuum energy to be GUT scale, $\alpha \lambda \phi_0^6 \approx (10^{-6}m_{pl})^4$ and $\epsilon = 1/30$. We include a homogeneous expanding background.

\begin{figure*}[tb] 
   \centering
   \includegraphics[width= 6.8in]{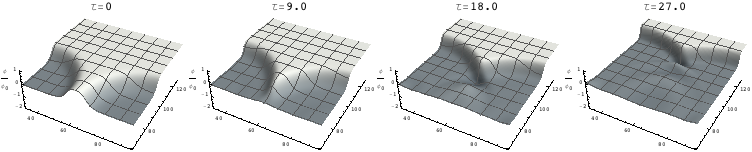} 
   \caption{Time evolution of two bubbles whose centers are $\sim .36 H_0^{-1}$ apart at nucleation.  When these bubbles collide (at $\tau\approx 9.0$) the walls have a Lorentz factor $\gamma\approx1.3$.  We use a conformal time $ a(t) d\tau =  \sqrt{\lambda}\phi_0^2 dt$, where $a(0)=1$ at the beginning of the simulation and $dt$ is the usual proper time of an FRW spacetime. Notice the Lorentz contraction of the wall thickness as the bubble expands.}
   \label{fig:notransition}
\end{figure*}

We begin with a case where the transition does not occur.  The two bubbles nucleate close together, at a separation $2.4R_0$.  In a static universe, the bubbles would collide when $r=1.2R_0$, but expansion of the background delays the collision until  $r=1.3R_0$ and $\gamma\approx1.3$.  Figure \ref{fig:notransition} shows the time evolution of this scenario.  The field does not have sufficient kinetic energy after the collision to climb over the potential barrier, and hence the bubbles merge into a single region with $\phi = \phi_2 \approx 0$.

Now consider two bubbles which nucleate at a separation of $3 R_0$. The bubble walls achieve $\gamma \approx 1.8$ before collision, and  then successfully traverse the potential barrier.  Figure \ref{fig:conform} shows a conformal diagram of the  resulting field profile, showing the classical nucleation of a bubble with vacuum energy $V(\phi_3)$.
We ran several simulations, varying the distance between nucleation points. Generically, we find that the critical Lorentz factor at which a transition occurs is $\gamma \approx 1.4$, or $\beta \approx 0.5$ via Eqn. (\ref{eqn:condition2}). There are several interesting features. First, classical transitions occur even when the bubbles are only mildly relativistic. Second, if transitions do not occur, \emph{most} of the energy is then released as debris (radiation) after the collision.
Third, the collision induced bubble has a very small initial size (of the order of wall thickness)
and is nucleated with a non-stationary wall.
The bubble is homogeneous inside and initially non-spherically symmetric, though later expansion tends to make
it more spherical.
Fourth, our 3D simulations confirm the robustness
of the hyperbolic symmetry: grossly unstable symmetry violating perturbations are not seen.

\begin{figure*}[tb] 
   \centering
   \includegraphics[width= 6.8in]{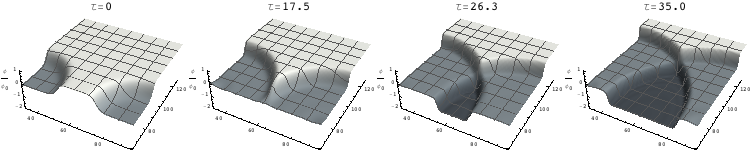} 
   \caption{Time evolution of two bubbles whose centers are $\sim .45 H_0^{-1}$ apart at nucleation.  When these bubbles collide (at $\tau\approx 17.5$) the walls have a Lorentz factor $\gamma=1.8$.  We use a conformal time $ a(t) d\tau =  \sqrt{\lambda}\phi_0^2 dt$, where $a(0)=1$ at the beginning of the simulation and $dt$ is the usual proper time of an FRW spacetime.}
   \label{fig:example}
\end{figure*}

{\em Discussion:} We have shown that, for a single scalar field model with several minima, bubble collisions generically lead to the formation of new, lower vacuum bubble unless the orignal bubbles nucleate very close to one other. 
This mechanism creates a new set of possibilities for old problems.

\emph{1. Are classical transitions the dominant bubble formation process?} 
Consider again the setup in Fig. \ref{fig:potential}. 
Let the nucleation rate from $1$ to $2$ be $\Gamma_{12}$, and that from $2$ to $3$ be $\Gamma_{23}$.
If $\Gamma_{12}$ is too high, $\phi_2$-bubbles will form too close to each other, and hence their wall energies will
be too small for them to classically transition to 
$\phi_3$-bubbles during collisions.  In this case the universe percolates rapidly into the $\phi_2$ vacuum, which
eventually quantum nucleates $\phi_3$-bubbles. For this not to happen, the $\phi_2$-bubbles need to
be nucleated at a typical separation, $\sim \Gamma_{12}^{-1/4}$, that is larger than
the initial bubble radius $R_0$ by a minimum Lorentz factor of $\gamma_{\rm min.}$ 
($\gamma_{\rm min.}^2$ is defined by the right hand side of Eqn. \ref{eqn:condition2}).
Therefore, a necessary condition for classical transitions to be important is
\begin{eqnarray}
\Gamma_{12} \lsim (\gamma_{\rm min.} R_0)^{-4} \label{eqn:dominate}
\end{eqnarray}
Whether classical transitions dominate over quantum tunneling into $\phi_3$
depends further on $\Gamma_{23}$. The question involves complex measure issues:
given a random point in the $\phi_3$ vacuum, what is the probability that it arose
from quantum tunneling or classical collisions? We will address this question elsewhere.

\emph{2. Implications for the eternally inflating stringy landscape.} 
Collisions, if they are sufficiently relativistic, provide a new way of scanning an eternally inflating
landscape. A collision could allow classical transitions not just over one or two barriers but multiple barriers.
It is thus necessary to revisit predictions for bubble counting measures based on quantum tunneling alone
\cite{Easther:2005wi,Garriga:2005av}.
We caution however that
multidimensional potentials generically possess complicated intra-field couplings, and such couplings can change the 
dynamics of the collision. For example, consider our toy model with an additional coupling to a light field $\chi$. 
This light field can be excited in a collision, and carry away energy thus increasing $\beta$, making the collision 
less elastic.

\emph{3. Observational signatures?} It is conceivable the bubble which is our own universe is formed via
a classical transition of the type discussed here. Our simulations suggest the
bubble is quite homogeneous but is initially highly anisotropic. Is the subsequent expansion sufficient
to make it acceptably isotropic? Or do we need some period of slow-roll inflation to make it both
isotropic and flat? Is the residual anisotropy observable? The
observational signatures are likely different from those considered by
\cite{Garriga:2006hw,Chang:2007eq,Aguirre:2007an}.

\emph{4. Implications for the small cosmological constant problem.} Abbott \cite{Abbott:1984qf} (see also
\cite{Brown:1987dd}) proposed a model for relaxing the cosmological constant using a step-wise potential 
via a series of tunneling events. Collisions introduce new, and perhaps faster, excursions through 
the multiple descending vacua. Tapping the collision energy might even help alleviate the well known
empty universe problem \cite{Guth:1982pn}.

To summarize, we presented a new mechanism of bubble nucleation where a potential barrier (or even multiple barriers) 
can be transitioned classically via energy released during collisions of two or more bubbles. This new mechanism possesses a rich phenomenology which we will explore, along with improved analytic descriptions of
the bubble formation criteria, in forthcoming publications.

\mbox{}

{\em Acknowledgments:}  We thank Matthew Johnson, Matthew Kleban, Adam Brown, Puneet Batra and Erick Weinberg for useful discussions.  Research at the Perimeter Institute for Theoretical Physics is supported by the Government of Canada through Industry Canada and by the Province of Ontario through the Ministry of Research \& Innovation.  RE is supported by the Department of Energy (DE-FG02-92ER-40704), the FQXi (RFP1-06-17) and  the NSF (CAREER-PHY-0747868). EAL is partly supported by a Foundational Questions Institute (FQXi) Mini-Grant.
LH is supported by the DOE (DE-FG02-92-ER40699) and the RISE program at Columbia.


\begin{thebibliography}{10}
\bibitem{Coleman:1980aw}
  S.~R.~Coleman and F.~De Luccia,
Phys.\ Rev.\  D {\bf 21}, 3305 (1980).

\bibitem{Hawking:1982ga}
S.~W.~Hawking, I.~G.~Moss and J.~M.~Stewart,
Phys.\ Rev.\  D {\bf 26} (1982) 2681.

\bibitem{Kosowsky:1991ua}
  A.~Kosowsky, M.~S.~Turner and R.~Watkins,
      Phys.\ Rev.\  D {\bf 45}, 4514 (1992).

\bibitem{Aguirre:2009}
  A.~Aguirre, M.~C.~Johnson and M.~Tysanner,
  Phys.\ Rev.\  D {\bf 79}, 123514 (2009)
  [arXiv:0811.0866 [hep-th]].

\bibitem{Langlois:2001uq}
  D.~Langlois, K.~i.~Maeda and D.~Wands,
        Phys.\ Rev.\ Lett.\  {\bf 88}, 181301 (2002)
          [arXiv:gr-qc/0111013].

\bibitem{Chang:2007eq}
S.~Chang, M.~Kleban and T.~S.~Levi,
JCAP {\bf 0804}, 034 (2008)
[arXiv:0712.2261 [hep-th]].

\bibitem{Aguirre:2007an}
  A.~Aguirre, M.~C.~Johnson and A.~Shomer,
  Phys.\ Rev.\  D {\bf 76}, 063509 (2007)
  [arXiv:0704.3473 [hep-th]].

\bibitem{Aguirre:2007anwm}
  A.~Aguirre and M.~C.~Johnson,
  Phys.\ Rev.\  D {\bf 77}, 123536 (2008)
  [arXiv:0712.3038 [hep-th]].

\bibitem{Felder:2000hq}
  G.~N.~Felder and I.~Tkachev,
  [arXiv:hep-ph/0011159].

\bibitem{Easther:2005wi}
  R.~Easther, E.~A.~Lim and M.~R.~Martin,
            JCAP {\bf 0603}, 016 (2006)
                    [arXiv:astro-ph/0511233].

\bibitem{Garriga:2005av}
J.~Garriga, D.~Schwartz-Perlov, A.~Vilenkin and S.~Winitzki,
JCAP {\bf 0601}, 017 (2006)
[arXiv:hep-th/0509184].

\bibitem{Garriga:2006hw}
  J.~Garriga, A.~H.~Guth and A.~Vilenkin,
  Phys.\ Rev.\  D {\bf 76}, 123512 (2007)
  [arXiv:hep-th/0612242].



\bibitem{Abbott:1984qf}
  L.~F.~Abbott,
      Phys.\ Lett.\  B {\bf 150}, 427 (1985).

\bibitem{Brown:1987dd}
  J.~D.~Brown and C.~Teitelboim,
      Phys.\ Lett.\  B {\bf 195}, 177 (1987).

\bibitem{Guth:1982pn}
  A.~H.~Guth and E.~J.~Weinberg,
        Nucl.\ Phys.\  B {\bf 212}, 321 (1983).

\end{thebibliography}
\end{document}